\input psfig
\documentstyle[12pt]{article}

\def\gtap{\raisebox{-.55ex}{\rlap{$\sim$}} \raisebox{.4ex}{$>$}}
\def\gsim{\mathrel{\gtap}}

\def\e{\mbox{e}}

\def\half{{1 \over 2}}
\def\a{\alpha}
\def\b{\beta}
\def\k{\kappa}
\def\t{\Theta}

\catcode`\@=11
\long\def\@makefntext#1{
\protect\noindent \hbox to 3.2pt {\hskip-.9pt
$^{{\ninerm\@thefnmark}}$\hfil}#1\hfill}		

\def\@makefnmark{\hbox to 0pt{$^{\@thefnmark}$\hss}}  
	
\def\ps@myheadings{\let\@mkboth\@gobbletwo
\def\@oddhead{\hbox{}
\rightmark\hfil\ninerm\thepage}
\def\@oddfoot{}\def\@evenhead{\ninerm\thepage\hfil
\leftmark\hbox{}}\def\@evenfoot{}
\def\sectionmark##1{}\def\subsectionmark##1{}}

\renewcommand{\thefootnote}{\fnsymbol{footnote}}

\newcounter{sectionc}\newcounter{subsectionc}\newcounter{subsubsectionc}
\renewcommand{\section}[1] {\vspace*{0.6cm}\addtocounter{sectionc}{1}
\setcounter{subsectionc}{0}\setcounter{subsubsectionc}{0}\noindent
	{\normalsize\bf\thesectionc. #1}\par\vspace*{0.4cm}}
\renewcommand{\subsection}[1] {\vspace*{0.6cm}\addtocounter{subsectionc}{1}
	\setcounter{subsubsectionc}{0}\noindent
	{\normalsize\it\thesectionc.\thesubsectionc. #1}\par\vspace*{0.4cm}}
\renewcommand{\subsubsection}[1]
{\vspace*{0.6cm}\addtocounter{subsubsectionc}{1}
	\noindent {\normalsize\rm\thesectionc.\thesubsectionc.\thesubsubsectionc.
	#1}\par\vspace*{0.4cm}}

\newcounter{appendixc}
\newcounter{subappendixc}[appendixc]
\newcounter{subsubappendixc}[subappendixc]

\renewcommand{\appendix}[1] {\vspace*{0.6cm}
        \refstepcounter{appendixc}
        \setcounter{figure}{0}
        \setcounter{table}{0}
        \setcounter{equation}{0}
        \renewcommand{\thefigure}{\Alph{appendixc}.\arabic{figure}}
        \renewcommand{\thetable}{\Alph{appendixc}.\arabic{table}}
        \renewcommand{\theappendixc}{\Alph{appendixc}}
        \renewcommand{\theequation}{\Alph{appendixc}.\arabic{equation}}
        \noindent{\bf Appendix \theappendixc #1}\par\vspace*{0.4cm}}

\def\abstracts#1{{
	\centering{\begin{minipage}{12.2truecm}\footnotesize\baselineskip=12pt\noindent
	\centerline{\footnotesize ABSTRACT}\vspace*{0.3cm}
	\parindent=0pt #1
	\end{minipage}}\par}}

\renewenvironment{thebibliography}[1]
	{\begin{list}{\arabic{enumi}.}
	{\usecounter{enumi}\setlength{\parsep}{0pt}
\setlength{\leftmargin 1.25cm}{\rightmargin 0pt}
	 \setlength{\itemsep}{0pt} \settowidth
	{\labelwidth}{#1.}\sloppy}}{\end{list}}

\newcounter{itemlistc}
\newcounter{romanlistc}
\newcounter{alphlistc}
\newcounter{arabiclistc}

\newcommand{\fcaption}[1]{
        \refstepcounter{figure}
        \setbox\@tempboxa = \hbox{\footnotesize Fig.~\thefigure. #1}
        \ifdim \wd\@tempboxa > 6in
           {\begin{center}
        \parbox{6in}{\footnotesize\baselineskip=12pt Fig.~\thefigure. #1}
            \end{center}}
        \else
             {\begin{center}
             {\footnotesize Fig.~\thefigure. #1}
              \end{center}}
        \fi}

\newcommand{\tcaption}[1]{
        \refstepcounter{table}
        \setbox\@tempboxa = \hbox{\footnotesize Table~\thetable. #1}
        \ifdim \wd\@tempboxa > 6in
           {\begin{center}
        \parbox{6in}{\footnotesize\baselineskip=12pt Table~\thetable. #1}
            \end{center}}
        \else
             {\begin{center}
             {\footnotesize Table~\thetable. #1}
              \end{center}}
        \fi}

\def\@citex[#1]#2{\if@filesw\immediate\write\@auxout
	{\string\citation{#2}}\fi
\def\@citea{}\@cite{\@for\@citeb:=#2\do
	{\@citea\def\@citea{,}\@ifundefined
	{b@\@citeb}{{\bf ?}\@warning
	{Citation `\@citeb' on page \thepage \space undefined}}
	{\csname b@\@citeb\endcsname}}}{#1}}

\newif\if@cghi
\def\cite{\@cghitrue\@ifnextchar [{\@tempswatrue
	\@citex}{\@tempswafalse\@citex[]}}
\def\citelow{\@cghifalse\@ifnextchar [{\@tempswatrue
	\@citex}{\@tempswafalse\@citex[]}}
\def\@cite#1#2{{$\null^{#1}$\if@tempswa\typeout
	{IJCGA warning: optional citation argument
	ignored: `#2'} \fi}}

 1
 1
 1

\font\ninerm=cmr9

\begin{document}

\centerline{\normalsize\bf METASTABLE DEFECTS IN GENERIC EXTENSIONS}
\centerline{\normalsize\bf OF THE STANDARD MODEL
\footnote{To
appear in the Proceedings of the Conference: QUARKS-96,
Yaroslavl, Russia; May 5-11, 1996. Institute for Nuclear Research, 1996.}
}
\baselineskip=15pt
 
\vspace*{0.6cm}
\centerline{\footnotesize T. N. TOMARAS 
}
\baselineskip=13pt
\centerline{\footnotesize\it Department of Physics and Institute of Plasma 
Physics, }
\baselineskip=12pt
\centerline{\footnotesize\it University of Crete, }
\baselineskip=12pt
\centerline{\footnotesize\it and }
\baselineskip=12pt
\centerline{\footnotesize\it Research Center of Crete} 
\baselineskip=12pt
\centerline{\footnotesize\it P.O. Box 2208, 710 03 Heraklion, Crete, Greece}
\centerline{\footnotesize E-mail: tomaras@physics.uch.gr}
\vspace*{0.9cm}
\abstracts{The main features of the recently found 
classically stable quasi-topological 
membranes and strings in generic, 
topologically trivial and weakly coupled two-Higgs extensions of the 
standard model are briefly reviewed.  
A variety of localized solutions 
in the same model are also presented.
}


\section{Introduction}
\indent
It is quite conceivable that the Higgs sector of 
the effective theory of electroweak
interactions will prove to contain more multiplets than just the
one doublet of the Minimal Standard Model (SM).
In fact there are well known theoretical arguments 
why this might even be desirable\cite{vivlio}.
Supersymmetry is one of them. The Minimal Supersymmetric Standard Model (MSSM)
contains two Higgs doublets. Models derived from superstrings 
as low energy effective theories also have generically a richer
Higgs sector. Finally, if nature employs the mechanism
of electroweak baryogenesis to produce enough 
$n_B/n_\gamma$ in the Universe\cite{baryogenesis}, then again 
an extended Higgs sector is necessary as an extra source of soft 
CP-violation. 

Let us consider such a model.
Assume in addition that it has  
topologically trivial target space and vacuum manifold, so that
it does not support the existence of
any kind of absolutely stable 
topological solitons\footnote{In the
opposite case topological solitons 
will exist in addition to the defects
discussed below.}. 
It was pointed out recently\cite{ribbons} that it is possible, 
if the Higgs masses satisfy certain lower bounds, 
to obtain a ${\it dynamical\; exclusion}$ 
of part of the target space, which 
becomes effectively non-trivial
and leads to the existence of ${\it metastable}$ quasi-topological
solitons, whose life-times depend on the specific values of the masses
and are typically cosmological. 
More importantly it was shown that the necessary conditions
may be satisfied for perturbatively small values of the coupling constants,
consistent with the semiclassical approximation.

In this lecture I will first
describe the method used to search for such solitons in the context
of a toy model\cite{ribbons} \cite{pallis}, 
and then I will apply it to find classically stable
wall\cite{membranes} and string\cite{strings} 
defects in the weakly-coupled 
generic 2HSM. 
This is one potentially 
interesting realization of a general result
which should be taken into account in the analysis of 
field theoretic models.

\setcounter{footnote}{0}
\renewcommand{\thefootnote}{\alph{footnote}}

\section{ A toy model }

The simplest way to present the method we use 
to search for classically stable 
quasi-topological solitons in a field theory, is in the context of 
the toy model\cite{ribbons}
$$ 
{\cal L} = {1\over 2} \partial_\mu \Phi^* \partial^\mu \Phi - 
{\lambda\over 4} (\Phi^* \Phi - v^2)^2 + \mu^2 v Re(\Phi)
\eqno(1)
$$
in 1+1 dimensions. First of all, the model has a unique vacuum and a trivial 
target space.
Thus, it does not have any topological solitons. 
Notice on the other hand, that in the limit $\lambda \rightarrow \infty $ 
all finite energy 
configurations have the form $\Phi = v \exp{i \Theta}$. The dynamics of 
the angle $\Theta$ is obtained by substituting this $\Phi$ into 
the action. One obtains the sine-Gordon model, 
$$ 
{\cal L} \rightarrow v^2 \Bigl[ {1\over 2} (\partial_\mu \Theta)^2 + 
\mu^2 cos\Theta \Bigr]
\eqno(2)
$$
known to possess absolutely stable topological
solitons, in which the angle $\Theta({\rm x})$ changes by $2 \pi$ as one
moves from $-\infty$ to $+\infty$.
It is then intuitively natural to expect that even for finite values of
$\lambda$, once it is kept large enough, remnants of these solitons will
survive not as absolutely stable solitons, since the topology of the model
will be lost, but as ${\it local\; minima}$ of the energy functional. Roughly,
the only way an initial configuration with non-zero winding can lose its 
topological charge and evolve into radiation around the trivial vacuum, is
when the Higgs magnitude passes through zero, a process strongly supressed
by a large radial-Higgs mass, i.e. by a large ratio 
$
R \equiv m/\mu \simeq \sqrt{2 \lambda} v / \mu.
$ 
Indeed, we have checked numerically\cite{pallis} 
that for $R \geq 6.1$ the model possesses
non-trivial local minima of the energy. These are static solutions of the 
field equations, characterized by non-zero winding of the phase $\Theta$ 
(like the topological solitons of the limiting theory), and a space
dependent Higgs magnitude, smaller that its asymptotic vacuum value 
inside the region of width $\sim 1/\mu$ of $\Theta$ variation.

A couple of comments are in order at this point: (a) It is
straightforward to check that the solitons exist for arbitrary values of the
parameter $\lambda$. Something very desirable since our semiclassical
discussion cannot be trusted outside the perturbative regime of small $\lambda$.
Indeed, one may by
appropriate rescalings pull outside of the action the factor $1/\lambda$ 
and the only physically relevant parameter left inside is the ratio R defined
above. Thus, all conditions about existence and stability of classical
solutions are conditions on R, while the quantum decay rate is 
exponentially supressed and of ${\cal O}(\exp(-1/\lambda))$ for small
$\lambda$\cite{ribbons}.
(b) All results are consistent with perturbative unitarity.
(c) One may analytically verify these results  
in a space-compactified version of the model.

\section{ The two-Higgs standard model } 

We will now use the above method to search for analogous
solitons in realistic two-Higgs-doublet extensions of the standard model.

The generic 2HSM is described by the Lagrangian 
$$ {\cal L} = 
  -{1\over 4} W^a_{\mu\nu} W^{a\mu\nu} -{1\over 4} Y_{\mu\nu} 
Y^{\mu\nu} 
+ \vert D_\mu H_1\vert^2 + \vert D_\mu H_2\vert^2
-V(H_1,H_2) \  
\eqno(3)
$$
where 
 $ W^a_{\mu\nu}=\partial_\mu W^a_\nu - \partial_\nu W^a_\mu - g \epsilon^
{abc} W^b_\mu W^c_\nu $ and  
$ Y_{\mu\nu}=\partial_\mu Y_\nu - \partial_\nu Y_\mu 
 $, the physical $Z^0$ and photon fields are
$Z_\mu =  W^3_\mu {\rm cos}\theta_W - Y_\mu {\rm  sin}\theta_W$ and
$A_\mu =  W^3_\mu {\rm sin}\theta_W + Y_\mu {\rm  cos}\theta_W$
and 
${\rm tan} \theta_W = g^\prime/ g$. 
Both Higgs doublets have hypercharge equal to one, 
the covariant derivative is 
$$ 
D_\mu H_I=(\partial_\mu + {i\over2} g \tau^a W^a_\mu + {i\over2} g^\prime 
Y_\mu) H_I 
$$ 
for $I=1,2$ and the potential reads
$$
 V(H_1,H_2) = \lambda_1 \Bigl( \vert H_1\vert^2 - {v_1^2\over2}\Bigr)^2 +
 \lambda_2 \Bigl( \vert H_2\vert^2 - {v_2^2\over 2}\Bigr)^2 +
\lambda_3 \Bigl( \vert H_1\vert^2  + \vert H_2\vert^2 - 
{v_1^2+v_2^2\over2}\Bigr)^2
$$
$$
+ \lambda_4 \Bigl[ \vert H_1\vert^2 \vert H_2\vert^2 - (H_1^\dagger
H_2)  (H_2^\dagger H_1)\Bigr] 
+\lambda_5 \Bigl[{\rm Re }(H_1^\dagger H_2) -{v_1 v_2\over2} \cos\xi\Bigr]^2 
$$
$$
+\lambda_6 \Bigl[ {\rm Im}( H_1^\dagger H_2)-{v_1 v_2\over2} \sin\xi\Bigr]^2 
\eqno(4)
$$
where $\vert H_I \vert^2 \equiv  H_I^\dagger H_I$. 
This is the most general potential\cite{vivlio}
subject to the condition that
both CP invariance and a discrete $Z_2$ symmetry ($H_1\rightarrow -H_1$)
are only broken softly. The softly broken $Z_2$ symmetry is there
to suppress unacceptably large flavor-changing neutral currents.  
Assuming all the $\lambda_i$ are positive, the minimum of the 
potential is at
$$
<H_1> =  \  e^{-i\xi} \left(  \matrix{ 0 \cr  v_1/ \sqrt{2} \cr} \right)
\ \ {\rm and} \ \ 
<H_2> =  \left( \matrix{ 0 \cr { v_2/ \sqrt{2}}\cr } \right) \ .
\eqno(5)
$$
and apart from the electroweak gauge bosons
with masses $m_\gamma = 0$, $m_W^2= g^2 (v_1^2+v_2^2)/4$ and 
$m_Z = m_W/\cos\theta_W$ , 
the perturbative spectrum contains a charged Higgs boson $H^+$,
a CP-odd neutral scalar $A^0$,
and two CP-even neutral scalars $h^0$ and $H^0$.
Again, like in the toy model, the Higgs vacuum is unique
and the target space topologically trivial (${\cal R}^8$), 
so that no topological solitons arise.

\subsection{ The membranes\cite{membranes} } 

Let us reduce the large number of parameters
by restricting ourselves to\footnote{Relaxing these conditions is 
straightforward but beyond the scope of the present note.}
$\lambda_1=\lambda_2$, $\lambda_5=\lambda_6$, $\xi = 0$ and
$v_1=v_2=v$ (or ${\rm tan}\beta = 1$) 
and consider the limit $\;\lambda_1=\lambda_2 \rightarrow \infty$ and 
$\lambda_4 \rightarrow \infty$. The first fixes the magnitudes of the two
doublets, while the second restricts them to the form \footnote{We made use of 
the gauge freedom of the model to assign half of the relative 
phase $\Theta$ between the two doublets on each one of them.}
$$H_1 = e^{i\Theta/2} \left(  \matrix{ 0 \cr  v/ \sqrt{2} \cr} \right)
\ \ {\rm and} \ \ 
H_2 = e^{-i\Theta/2}  \left(  \matrix{ 0 \cr  v/ \sqrt{2} \cr} \right)\ .
\eqno(6)
$$
In the limit we are considering only $A^0$ has finite mass, so only the 
corresponding field $\Theta$ can have non-trivial space
dependence in finite energy configurations.

The energy functional for the field $\Theta$ takes the sine-Gordon form
$$ 
E  = {v^2\over 2} \int d^3x \Bigl[ {1\over 2} (\nabla \Theta)^2 - 
m_A^2 \cos\Theta \Bigr] + {\rm gauge}
\eqno(7)
$$ 
where the gauge contribution is minimum when $W_\mu^a = Y_\mu = 0$.
The limiting sine-Gordon model possesses absolutely stable topological walls,
of thickness of ${\cal O}(m^{-1}_A)$ and energy per unit area 
$
{\cal E}/{\cal A} = 4 v^2 m_A = 2 m_W^2 m_A {\rm sin}^2\theta_W / \pi \alpha
$
($\alpha$ is the fine structure constant). We thus expect that for finite
but "large enough values of the parameters 
$\lambda_1=\lambda_2$ and $\lambda_4$" there will arise classically stable
solutions of the full equations of the model which should look very
similar to the sine-Gordon solution. 

The search for solutions and the study of their stability for finite values 
of the parameters $\lambda_1 = \lambda_2$ and $\lambda_4$ was carried out 
numerically in the gauge $W^a_x = Y_x = 0$. The profile of the solution 
for $m_h = 2.5 m_A, m_{H^0} = 5.0 m_A$ and $m_{H^+} = 4.0 m_A$ is shown in
Figure 1. 

\hbox{
\hspace{0.57cm}
\psfig{file=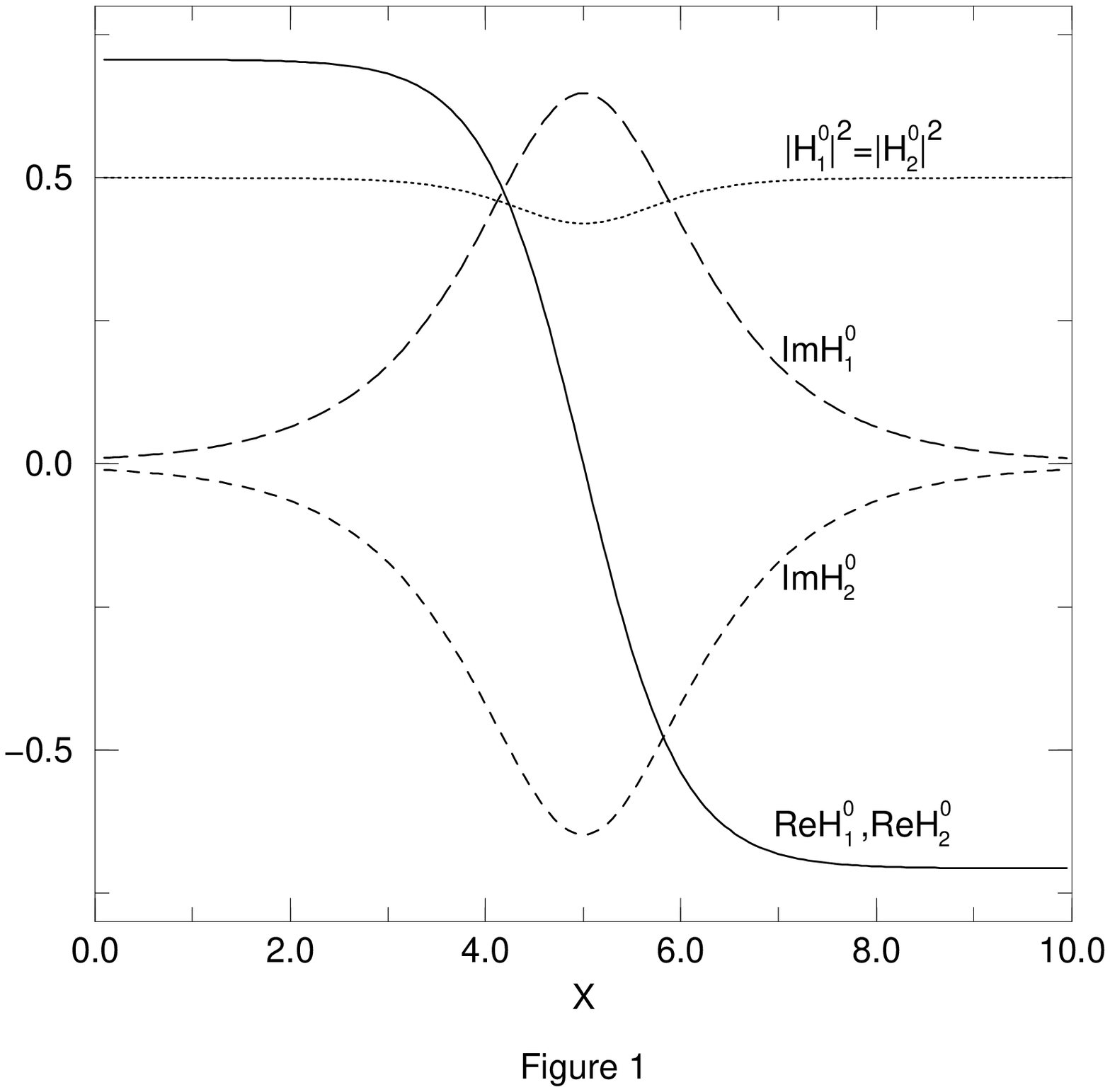,bbllx=50bp,bblly=125bp,bburx=500bp,bbury=620bp,width=10.5cm,angle=0}
}

All gauge fields as well as the upper (charged) components of
both Higgs doublets vanish. The neutral components of the doublets differ 
very little from their form in the limiting theory. This fact is also shown
in Figure 2, where we plot the neutral components of the two Higgs doublets
in the complex plane. 
Their phases wind around  in opposite
directions as one crosses the wall, starting at zero and joining
at $\pm\pi$.

Figure 3 depicts the boundaries of classical stability region in the
$(m_{H^0}, m_{H^+})$ plane for three different values of $m_h$ all in units
of $m_A$.
Classically-stable membranes exist above the indicated lines.
Also given is the energy density in units of $v^2 m_A$, for some
selected points close to the boundaries.
For values of the Higgs masses consistent with present experimental bounds and
with perturbative unitarity (for instance for $m_A \sim$50 GeV, 
$m_h \sim$125 GeV and $m_{H^0}$, $m_{H^+} \gsim$200 GeV) the theory
supports the existence of classically stable wall "defects", with typical
thickness and energy density of ${\cal O}(m_A^{-1})$ and 
${\cal O}(4 m_W^3) \sim 10^{10} gr/cm^2$
respectively. Their decay rate 
per unit area has an exponential dependence on the parameters of the theory
and is thus very sensitive to their precise values.

\hbox{
\hspace{0.57cm}
\psfig{file=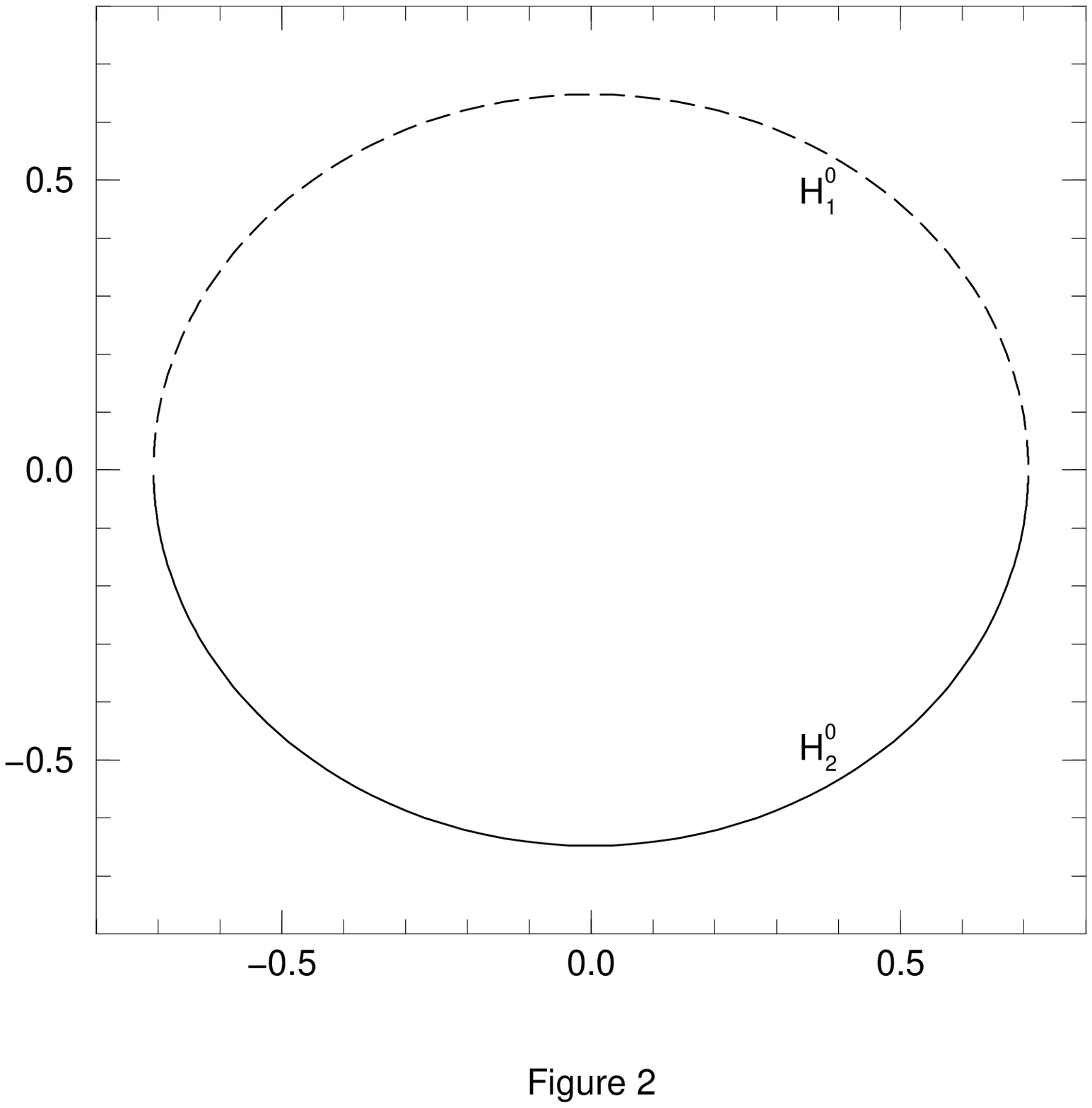,bbllx=50bp,bblly=125bp,bburx=500bp,bbury=620bp,width=10.5cm,angle=0}
}

They are so heavy that if their life-times are cosmological 
they will be as desastrous as the ordinary 
topological domain walls arising in models with spontaneously broken
discrete symmetries. The presence of even an isolated wall of
cosmic dimensions 
in our Universe today is clearly excluded. 
Assuming, as it seems quite natural,
that their production in the early universe is not extremely suppressed, 
the curves of Figure 3 should in this case be interpreted as giving 
the ${\it upper}$ bounds of the Higgs masses consistent with the 
cosmological data.
If, on the other hand, the membranes are short lived (as it will happen if
the values of the masses happen to be close to a boundary line), 
they are potentially
useful either as initial seeds for galaxy formation, or by contributing to 
the electroweak baryogenesis. For these statements to become more 
quantitative a thorough study is necessary of their production mechanism, 
of their decay rate and of their dynamics.

\subsection{ The strings\cite{strings} }

It is possible to apply the same method to find stable 
string solutions in the 2HSM.
The details are given in Reference [6]. Here
I briefly present the main steps.

Let us go back to (3), (4) and consider the limit 
$\lambda_1, \lambda_2 \rightarrow \infty$ 
and $\lambda_5 \rightarrow \infty$. 
Among the Higgs fields only $H^+$ has finite mass in this limit.

\hbox{
\hspace{0.57cm}
\psfig{file=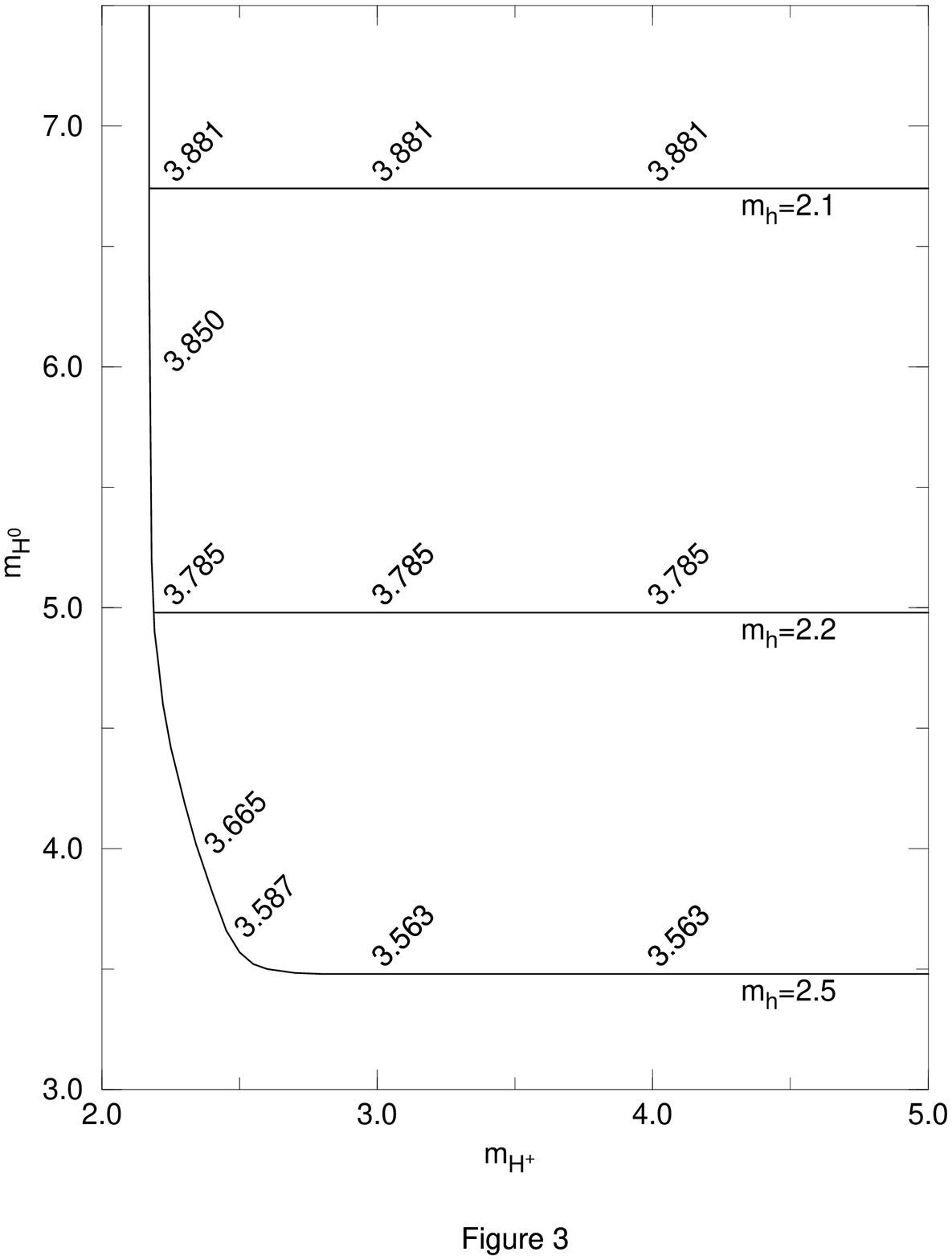,bbllx=50bp,bblly=40bp,bburx=500bp,bbury=690bp,width=11cm,angle=0}
}


The generic finite energy Higgs field configuration is then
$$H_1 = \left(  \matrix{ 0 \cr  v_1/ \sqrt{2} \cr} \right)
\ \ {\rm and} \ \ 
H_2 = U \left(  \matrix{ 0 \cr  v_2/ \sqrt{2} \cr} \right)\ ,
\eqno(8)
$$
with
$$
U(x) = e^{-i \,\xi\, {\bf n}(x)\, \cdot\, {\bf \tau}}
\eqno(9)
$$
The only dynamical field left in the above limit is the unit vector 
${\bf n}(x)$, which for finite energy must satisfy $n_3(x) = 1$ at spatial
infinity.

If in addition one takes the remaining couplings of the Higgs potential and the
gauge couplings to zero, one is left with an effective O(3) non-linear
$\sigma-$model for the unit vector field ${\bf n}(x)$. 
It is well known\cite{Belavin}
that this model possesses topological solitons characterized by an $S^2 \to
S^2$ winding number, an arbitrary size $\rho$ and an energy
which due to the scale invariance of the model does not depend on $\rho$.

What will happen to such a soliton solution of the limiting theory if one tries
to relax the couplings $\lambda_1$, $\lambda_2$ 
and $\lambda_5$ to finite values? To leading order one only has to consider
the effect of the potential and of the gauge 
interactions on the size of the soliton\cite{strings}. 
Using a scaling argument\cite{Derrick} one may show that the potential
tends to shrink the soliton and the question then is if the gauge interaction
is able to halt this shrinking. 
In Reference [6] it was shown that for large enough 
radial Higgs masses the two opposing forces come to equilibrium and a 
classically stable string solution arises. This is actually true even for
perturbatively small values of the parameters, again 
consistent with perturbative
unitarity and with present day phenomenological bounds on the Higgs masses. 

Such strings may be produced either cosmologically at the electroweak
phase transition or with smaller probability as finite length (of order 
a few over $m_W$)
string loops in large accelerators such as LHC. 
We do not have at this point control over their production
probability and their decay rate, necessary to analyse their
phenomenological implications. Naively, one expects 
that so energetic Higgs lumps will 
decay characteristically into a very large number of jets.

\subsection{No stable localized spherically symmetric solitons\cite{solitons}}

In a similar fashion, 
let us go back to the generic two-Higgs model, and set for simplicity 
$v_1=v_2\equiv v$ and  
$g^\prime, \lambda_3, \lambda_4, \lambda_5, \lambda_6 $ as well as $\xi$ 
all equal to zero. In the simplified model consider the limit
$\lambda_1 = \lambda_2 \rightarrow \infty$.  
All finite energy Higgs configurations must then have the form:
$$H_1 = \left(  \matrix{ 0 \cr  v/ \sqrt{2} \cr} \right)
\ \ {\rm and} \ \ 
H_2 = U \left(  \matrix{ 0 \cr  v/ \sqrt{2} \cr} \right)\ ,
\eqno(10)
$$
with $U$ a general SU(2) matrix. The dynamics of $U$ is described by
$$
{\cal L} \sim {\rm Tr} \partial_\mu U^{-1} \partial^\mu U + {\rm gauge\; part}
\eqno(11)
$$
One may follow the steps of the string case and consider the limit $g=0$.
The model reduces to the global SU(2) non-linear $\sigma$-model. Finite energy
configurations of this model have $U =$ constant at infinity and are thus
classified according to $\pi_3 (S^3)$. Contrary to the 
previous two-dimensional case
the limiting model does not support
the existence of such solutions. Any such non-trivial configuration
shrinks indefinitely to a singular configuration.

Our starting point is not so promising. In contrast to the previous
cases, the limiting theory does not possess stable solitons.
Actually, once we allow for finite values of $\lambda_1$ and $\lambda_2$
the shrinking tendency of any non-trivial Higgs configuration will
be enhanced. One might still hope though, that this will be halted
by the gauge interaction, in much the same way this happened
in the case of the strings.

The results of our search for spherically symmetric solutions and for
generic values of all parameters of the model are described in
detail in Reference [9]. 
There is a wide variety of solutions none of which was found to be stable.
Figure 4 is a graphical representation of the solution branches in the 
special case of equal masses $m_1 = m_2\equiv m$ for the two radial Higgses 
and equal values $v_1 = v_2$ 
for their corresponding vevs. 

\hbox{
\hspace{0.57cm}
\psfig{file=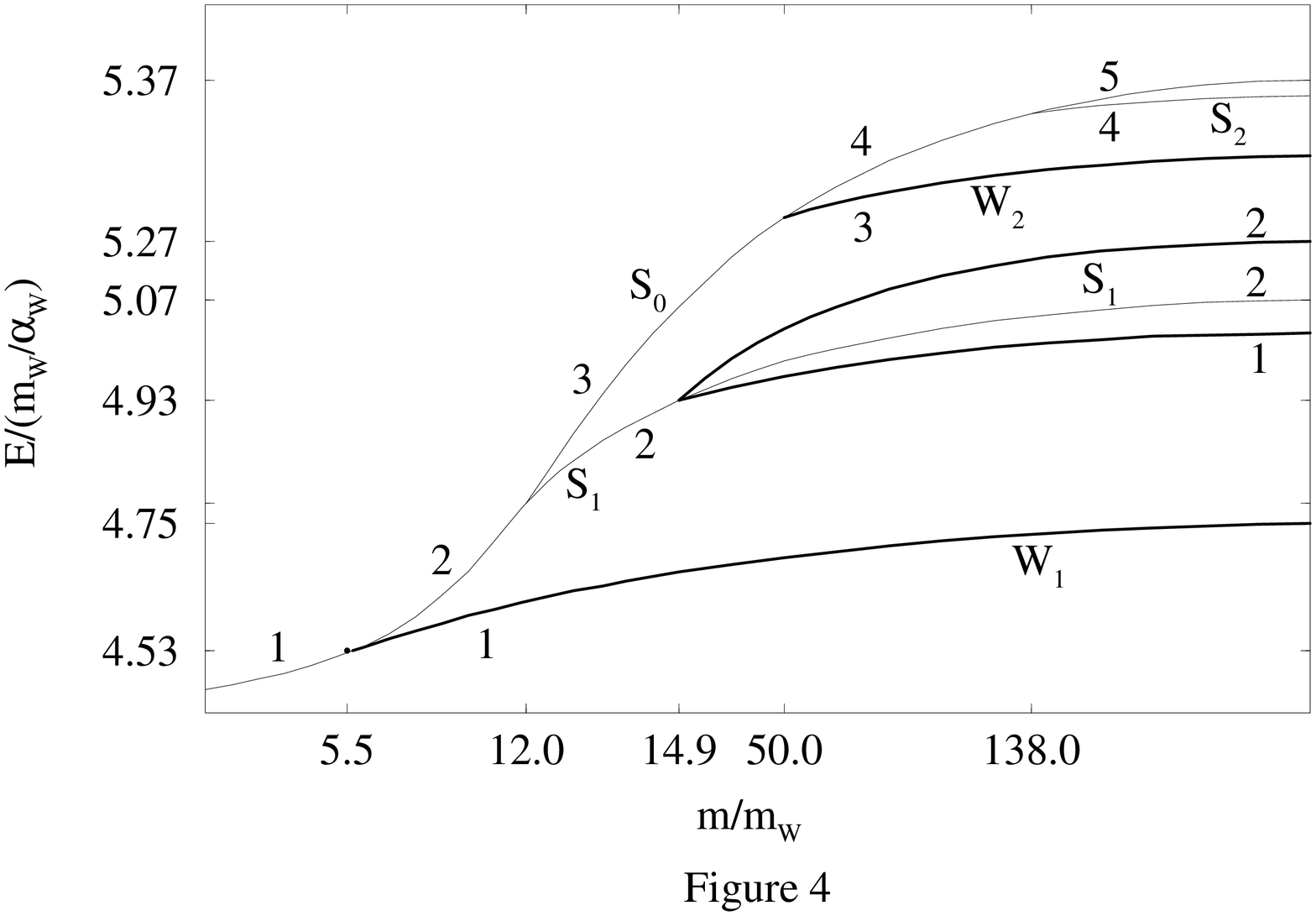,bbllx=50bp,bblly=90bp,bburx=500bp,bbury=790bp,height=10cm,width=14cm,angle=270}
}

\vspace {1.7cm}

$S_0$ corresponds to the sphaleron solution of the SM 
embedded in the 2HSM\cite{Peccei}, $S_1$ to
the first deformed sphaleron and so on. The $W_n$ are new branches with no 
analog in the SM. The integer number shown on each branch represents
the number of unstable modes of the corresponding solution. 
The interested reader may find a discussion of the characteristics 
of the various branches of solutions
and of their potential phenomenological role in Reference [9].

I would like to finish with a comment about the possibility of 
another kind of localized solitons in the 2HSM. It was argued above that there
exists a limit of the parameters of the model in which the only remaining
degrees of freedom are the components of a unit vector ${\bf n}$.
Furthermore, any finite energy configuration must have $n_3=1$ at 
spatial infinity. In three dimensions such a configuration defines 
a map from $S^3$ into $S^2$, known to be classified according to the 
Hopf homotopy
group\cite{Bott} $\pi_3(S^2)$. These maps are at best axially symmetric and can 
be thought of as a finite length straight 
string twisted by an integral multiple
of 2$\pi$ around its axis and glued at the two
ends. A situation in between the infinite string (shown to exist) and
the spherically symmetric case, where we failed. It is thus possible 
that in this case the gauge fields will manage to halt the shrinking 
and that non-trivial axially symmetric
stable Hopf configurations will exist. 

\vspace{1cm}
{\bf Acknowledgements}
\vskip 0.4cm  

I would like to thank Prof. S. Dimopoulos for a useful discussion about
the phenomenology of the defects discussed above.
This research was supported in part by
the EEC grants CHRX-CT94-0621 and CHRX-CT93-0340,
as well as by the Greek General Secretariat
of Research and Technology
grant 95$E\Delta$1759.

\vfill
\eject
\end{document}

\begin{thebibliography}{9}
\bibitem{DHN} R. Dashen, B. Hasslacher and A. Neveu, 
{\it Phys. Rev.} {\bf D10} (1974) 4138.

\bibitem{Taubes} C. Taubes, {\it Comm. Math. Phys.}
 {\bf 86} (1982) 257
and 299;\\
N.S. Manton, {\it Phys. Rev.} {\bf D28} (1983) 2019;\\
P. Forgacs and Z. Horvath, {\it Phys. Lett.}
 {\bf 138B} (1984) 397.

\bibitem{Klink} F. Klinkhamer and N. Manton, 
{\it Phys. Rev.} {\bf D30} (1984) 2212.

\bibitem{Boguta} J. Boguta, {\it Phys. Rev. Lett.}
 {\bf 50} (1983) 148;\\
J. Burzlaff, {\it Nucl. Phys.} {\bf B233} (1984) 262.

\bibitem{eila} G. Eilam, D. Klabucar and A. Stern, 
{\it Phys. Rev. Lett.} {\bf 56}
(1986) 1331;\\
G. Eilam and A. Stern, {\it Nucl. Phys.} {\bf B294} (1987) 775.

\bibitem{Brihaye} J. Kunz and Y. Brihaye, 
{\it Phys. Lett.} {\bf 216B} (1989) 353 and
 {\bf 249B} (1990) 90.

\bibitem{Yaffe} L. Yaffe, {\it Phys. Rev.} {\bf D40} (1989) 3463.

\bibitem{Klink2} F. Klinkhamer, {\it Phys. Lett.}
 {\bf 236B} (1990) 187.

\bibitem{Nambu} Y. Nambu, {\it Nucl. Phys.} {\bf B130} (1977) 505.

\bibitem{Soni} M.B. Einhorn  and R. Savit,
 {\it Phys. Lett.} {\bf 77B} (1978)
295;\\
 V. Soni, {\it Phys. Lett.} {\bf 93B} (1980) 101;\\
  K. Huang and R. Tipton, {\it Phys. Rev.} {\bf D23} (1981)
3050.

\bibitem{Vach} T. Vachaspati, {\it Phys. Rev. Lett.}
 {\bf 68} (1992) 1977\\
   and
{\it Nucl. Phys.} {\bf B397} (1993) 648;\\
T. Vachaspati and M. Barriola, {\it Phys. Rev. Lett.}
 {\bf 69} (1992) 1867.

\bibitem{Vach1} M. James, T. Vachaspati and L. Perivolaropoulos, 
{\it Phys. Rev.} {\bf D46} (1992)
R5232 and  {\it Nucl. Phys.} {\bf B395} (1993) 534.

\bibitem{Periv} L. Perivolaropoulos,
{\it Phys. Lett.} {\bf 316B} (1993) 528;\\
\ M. Earnshaw and M. James, {\it Phys. Rev.}
 {\bf D48} (1993) 5818.

\bibitem{Tze}J.M. Gipson and H.C.Tze,
 {\it Nucl. Phys.} {\bf B183} (1981) 524;\\
J.M. Gipson, {\it Nucl. Phys.} {\bf B231} (1984) 365.

\bibitem{Fahri} E. D'Hoker and E. Fahri,
{\it Nucl. Phys.} {\bf B241} (1984) 104, and \\
{\it Phys. Lett.} {\bf 134B} (1984) 86.

\bibitem{Rub} V.A. Rubakov, {\it Nucl. Phys.}
 {\bf B256} (1985) 509;\\
J. Ambjorn and V.A. Rubakov, {\it Nucl. Phys.}
 {\bf B256} (1985) 434.

\bibitem{Zee}  
 R. MacKenzie, F. Wilczek and A. Zee,
{\it Phys. Rev. Lett.} {\bf 53} (1984)
2203;\\
R. MacKenzie, {\it Mod. Phys. Lett.} {\bf A7} (1992) 293.

\bibitem{Carlson} J.W. Carlson,
{\it Nucl. Phys.} {\bf B253} (1985) 149, and 
{\bf B277} (1986) 253.

\bibitem{Skyrme} T.H.R. Skyrme,
{\it Proc. Roy. Soc.} (London), Ser. {\bf A260}
(1961) 127.

\bibitem{Derrick} G.H. Derrick, 
{\it J. Math. Phys.} {\bf 5} (1964) 1252.

\bibitem{Kunz} Y. Brihaye, J. Kunz and F. Mousset,
       {\it Z. Phys.} {\bf C56} (1992) 231;\\
J. Baacke, G. Eilam and H. Lange, {\it Phys. Lett.}
 {\bf 199B} (1987) 234.

\bibitem{ribbons} C. Bachas and T.N. Tomaras, {\it Nucl. Phys.} 
{\bf B428} (1994) 209.

\bibitem{pallis} K. Pallis, University of Thessaloniki
 undergraduate thesis (1994).

\bibitem{membranes} C. Bachas and T.N. Tomaras, hep-ph/9508395;
{\it Phys. Rev. Lett.} {\bf 76} (1996) 356.

\bibitem{strings} C. Bachas and T.N. Tomaras,
{\it Phys. Rev.} {\bf D51} (1995) R5356; C. Bachas, B. Rai and T.N. Tomaras,
Ecole Polytechnique and Crete preprint, to appear (September 1996).

\bibitem{solitons} C. Bachas, P. Tinyakov and T.N. Tomaras, "On spherically
symmetric solutions in the two-Higgs standard model", hep-ph/9606348; Accepted
for publication in Phys. Lett. B.

\bibitem{Giller} Y. Brihaye, S. Giller, P. Kosinski and J. Kunz,
{\it Phys. Lett.} {\bf 293B} (1992) 383.

\bibitem{Belavin} A.A. Belavin and A.M. Polyakov,
{\it JETP Lett.} {\bf 22} (1975) 245.

\bibitem{Zak} R.A. Leese, M. Peyrard and W.J. Zakrzewski,
{\it Nonlinearity} {\bf 3} (1990) 773.
 
\bibitem{Dvali} G. Dvali, Z. Tavartkiladze and J. Nanobashvili, {\it
Phys. Lett.} {\bf 352B} (1995) 214.

\bibitem{vivlio} J. Gunion, H. Haber, G. Kane and S. Dawson, {\it The
Higgs Hunter's Guide} (Frontiers in Physics, Addison-Wesley, 1990).
  
\bibitem{Kunzz}  Y. Brihaye, J. Kunz and
 C. Semay, {\it Phys. Rev.}
 {\bf D42} (1990) 193; \\
{\it Phys. Rev.} {\bf D44} (1991) 250,
and references therein.

\bibitem{Dobado}  
 A.Dobado and J.J. Herrero, {\it Nucl. Phys.}
{\bf B319} (1989) 491, 
and references therein.

\bibitem{Peccei} B. Kastening, R.D. Peccei and X. Zhang,
{\it Phys. Lett.} {\bf 266B} (1991) 413.

The next step is to show that in some limit of the parameters of the model
it possesses topological soliton solutions. 
In addition to the electroweak gauge bosons
with masses $m_W^2= g^2 v^2/2$ and $m_Z = m_W/cos\theta_W$ , 
the perturbative spectrum contains a charged Higgs boson $H^+$ with
mass $m_{H^+}^{\ 2} = \lambda_4 v^2 $ ,
a CP-odd neutral scalar $A^0$ with mass $m_{A}^{\ 2}= \lambda_5 v^2 $ ,
and two CP-even neutral scalars $h^0$ and $H^0$ with masses
$m_h^{\ 2} = 2\lambda_1 v^2$ and  
$m_{H^0}^{\ 2} = (2\lambda_1 + 4\lambda_3 + \lambda_5) v^2$ respectively.

The search for stable lumps in the Weinberg-Salam model
has a long history. 
It has revealed a rich
structure of classical solutions including the 
sphaleron\cite{DHN,Taubes,Klink,Boguta} ,
deformed sphalerons\cite{eila,Brihaye,Yaffe,Klink2}
 and vortex strings\cite{Nambu,Soni,Vach,Vach1,Periv}.
Such solutions could play a role in understanding
{\ninerm (B+L)}-violation
and structure formation in the early universe,
but they are all classically-unstable or/and extended.
They have therefore
no direct  present-day manifestation, 
contrary to long-lived particles whose
relic density could at least in principle be
detected.

The existence of particle-like excitations has, on the
other hand, been argued for in the context of a strongly-
interacting Higgs sector\cite{Tze,Fahri,Rub,Zee,Carlson}.
The advocated particles can be thought of as
technibaryons of
an underlying technicolor model.  
They are described in
bosonic language by winding solitons of
 an effective non-renormalizable
lagrangian for the   pseudo-goldstone-boson
 (or technipion) field,
much like  skyrmions\cite{Skyrme}
 of the effective chiral lagrangian
of $QCD$.  This is of course a phenomenological
 description, since
the   properties of such hypothetical 
particles cannot be
calculated reliably within a semi-classical expansion.
Furthermore, in 
view of the difficulties
facing technicolor models, 
the  possibility of a strongly-interacting Higgs sector is
not theoretically  appealing.

It would be clearly    more interesting
 if classically-stable winding excitations
could  arise in a {\it weakly-coupled} scalar
 sector.
To be more precise let us decompose the
 Higgs-doublet  field into a
real (positive)  magnitude   and a
group-phase: $\Phi = F U$, and
consider static configurations with $U(x)$
 wrapping $N$ times
around the ${\rm SU(2)}$ manifold. These are
 potentially unstable
for at least three distinct reasons:
{\it (a)} because $N$ is not   conserved 
whenever the magnitude $F$   goes through
zero;
{\it (b)} because $N$ is not gauge-invariant and can,
in particular, be non-vanishing
even in a vacuum state;  
and {\it (c)} because scalar-field configurations can loose 
their energy
by shrinking to zero size\cite{Derrick}.
We refer to these for short as the {\it radial, gauge}
 and {\it scale}
instabilities. They can be eliminated formally by
{\it (a)}
 taking the physical-Higgs mass $m_H\to\infty$, 
{\it (b)} decoupling the electroweak gauge fields,
 and 
{\it (c)} adding appropriate higher-derivative terms to the
 action.
The question is whether classical stability can
 be maintained
while relaxing the above conditions.
This
 has been investigated numerically in the 
{\it minimal} case of one  
doublet: although one may indeed relax both
the weak gauge coupling\cite{Rub,eila}
and  the Higgs mass\cite{Kunz}
  up to some finite critical values,    
  stability cannot apparently be achieved without the
   non-renormalizable  higher-derivative 
  terms in the
 action.
On the other hand, as we have 
 demonstrated recently,
  metastable winding solitons do arise in
renormalizable models in
two\cite{mexican} and three\cite{preprint}
space-time dimensions. The way this happens
is the content of section 2.
It is we believe instructive and could guide 
the search for lower dimensional structures in
many physical systems as well as for similar
semi-classical solitons in four dimensions.
Section 3 contains a brief
presentation of the status of our search for particle-like solutions
in the context of two-Higgs extensions of the standard model.
We close with some comments in the discussion section.

The simplest context in which the {\it radial} 
instability is an issue is a two-dimensional model of
a complex scalar field with mexican-hat potential:
$ V = {1\over 4}\lambda (\Phi^*\Phi - v^2)^2$. 
The simplest way to describe in detail the winding solitons 
is to take space to be periodic with period
$L$ \footnote{Alternatively we may add a mass term:
$\delta V = -\mu^2 v Re(\Phi)$,  that
lifts the vacuum degeneracy. Stable winding excitations, which reduce to
the sine-Gordon solitons in the $\lambda\to\infty$ limit,  
can be shown\cite{Pallis} numerically to exist for 
$\lambda v^2/\mu > 18.8$ \  .} . 
 The condition for classical stability can in this case
be derived analytically and reads\cite{mexican} : $m_H L > \sqrt{5} \ ,$
where $m_H=\sqrt{2\lambda} v$. The classically-relevant parameter
is thus the radial-Higgs mass in units of the soliton  size.
This follows also by comparing  the loss in potential energy to the gain
in gradient energy when trying to undo the winding by
reducing  the magnitude of the scalar. Note that
the loop-expansion parameter $\lambda L^2$, can be taken to zero 
independently so as to reach a semiclassical limit.

 The above winding solitons become unstable 
  classically if we gauge the $U(1)$ symmetry of the model.
The {\it gauge} instability is in fact
more severe than in four dimensions,
because   no energetic barrier 
  opposes the turning-on of
a static space-like gauge field, which is necessary
 to reach a winding-vacuum
state. The minimal abelian-Higgs model has thus only unstable
(sphaleron) solutions\cite{Giller}
 \footnote{It was claimed erroneously in [22] that it
has no static solutions whatsoever.
 This is only correct in the $\lambda\to\infty$
limit.}$\ $ .
The situation changes, however, drastically 
if there are more than one complex scalars. The
gauge-invariant  relative phases
of any two of them   cannot in this case  wind around
non-trivially in a vacuum state.
 An explicit  
analysis 
of this extended Abelian-Higgs model\cite{mexican}
 shows that   winding solitons
  persist down to scalar masses close to  
 the inverse soliton size: gauging and the extra Higgs  
  enhance the stability region found in the global model.

 The {\it scale} instability becomes an issue for
 the first time
in three space-time dimensions.
To be more precise we consider a real-triplet
 scalar field $\Phi_a(x)$ 
($a=1,2,3$) with mexican-hat
potential : $V= {1\over 4}\lambda (\Phi_a\Phi_a - v^2)^2$.
The  limit $\lambda\to\infty$ corresponds to the 
$O(3)$
non-linear $\sigma$-model. This is known to possess
 winding solitons, characterized by non-trivial
 mappings of the two-sphere
onto itself, and having
  arbitrary size\cite{Belavin}.
For finite $\lambda$ on the other hand, or in the
 presence of a
symmetry-breaking potential, Derrick's
 scaling argument\cite{Derrick}
 shows that these   solitons are unstable to shrinking.
One can of course again invoke higher-derivative terms 
to stabilize the scale\cite{Zak}.
The same result is however in this case  achieved by a massive
$U(1)$ gauge field with only renormalizable
 couplings\cite{preprint}.
This can be established
 by perturbing around the
$O(3)$ non-linear $\sigma$-model limit,
 or else by solving numerically
the equations of motion.

What do these lower-dimensional solitons teach us?
{\it First}, they suggest by analogy that 
classically-stable winding solitons may
exist in a weakly-coupled two-Higgs extension of the standard model.
The status of our search for such localized solutions is the content
of the following section.

{\it Second}, they are interesting in their own right, 
since they correspond
to a new class of wall and string defects in realistic
four-dimensional models of particle and condensed-matter 
physics . One such example are the metastable
{\it membranes}, discovered recently\cite{membranes,Dvali} in
two or more Higgs-doublet extensions of the standard model.
These are static
wall-type solutions, non-topological but classically stable 
in a wide region of the Higgs sector 
parameter space compatible with
perturbative unitarity and with present phenomenological bounds.
They are embeddings of the above discussed\cite{mexican}
solutions of the 2d Abelian-Higgs model with two or more complex scalars.
They are characterized by the non-trivial winding of
the relative U(1) phase of the neutral components of any
two Higgs doublets of the extended standard model in the
direction $x$ normal to the wall. They have no electromagnetic coupling, 
their size is a few times the inverse of the mass $m_A$ 
of the CP-odd scalar $A^0$ (in the standard notation of the generic
two-Higgs model\cite{vivlio}), while their energy per unit area
is in terms of the $A^0$ and W masses and the fine structure constant 
of order $m_W^2 m_A /\alpha$. 
Assuming a $m_A \simeq 50 GeV$ i.e. not much larger than
the present experimental lower bound\cite{vivlio}, the mass
of the wall is of ${\cal O} (10^{10} gr/cm^2)$. Thus, a single
wall crossing the entire Universe today would by far overclose it
and can be excluded. Smaller membranes though, which either
collapsed or were torn apart by quantum tunneling may have acted
as seeds for the formation of galaxies. In fact, the mass of
a typical galaxy is comparable to that of a membrane a few
light years in size. In any case, the production and decay
rates of these objects and their cosmological role 
has to be studied in detail, since most likely
they are going to lead to firm constaints on the Higgs mass
spectrum of extensions of the standard model\cite{membranes}.

As another example, we would like to mention the static 
wall-type metastable solutions of the easy-plane
ferromagnetic continuum in the presence of an external
in-plane magnetic field $h$ \footnote{The search for finite-energy
soliton solutions 
in ferromagnetic and antiferromagnetic systems and the study of their
stability, reported briefly in 
this paragraph,
was carried out in collaboration with Prof. N. Papanicolaou.
I would also like to thank Dr. P. Tinyakov for helpful discussions 
about sphaleron
solutions in relativistic anisotropic 
non-linear $\sigma-$models in 1+1-dimensions.} .
The dynamics of the system is described by the Landau-Lifchitz
equations with energy density
given in terms of the unit magnetization vector $n_a, \;a=1,2,3,$  
by:
$
{\cal E} = {1\over2} (\partial_i n_a)^2 +
g {n_3}^2 + h (1 - n_1).
$  
The anisotropy constant $g$ and the external field $h$ are 
both taken to be positive. 
The system has the unique semiclassical ground state: 
$(n_1=1, n_2=0=n_3)$ and
no topologically stable domain walls.
Nevertheless, it posesses a variety of wall-type static finite
energy (per unit area) solutions. A particularly interesting one is 
$(n_1=cos\Theta(x), n_2=sin\Theta(x), n_3=0)$, 
with $\Theta(x)$ satisfying $\Theta^{\prime \prime} (x) = h sin\Theta(x)$ 
and the boundary conditions $\Theta(-\infty)=0, \Theta(+\infty)=2 \pi$.
It can be verified that this solution is 
dynamically stable for $g/h \ge 3$.

\section{Localized solutions in 3+1 dimensions 
\footnote{The work in this section was
done in collaboration with Dr. P. Tinyakov and 
is described in detail in reference [31].} }

We now turn our attention to the possibility of classically stable
localized particle-like winding
solitons in 3+1 dimensions.
In the context of a two-Higgs extension of the standard model
these hypothetical solitons would: {\it (b)} be characterized
by the non-trivial winding
of the relative phase of the two doublets, and thus be immune to
the gauge mode of decay; {\it (c)} have a scale stabilized by 
electroweak magnetic fields and hence  of order $1/m_W$;
and {\it (a)}   hopefully stay stable for Higgs masses near $m_W$
and thus compatible with perturbative unitarity
 \footnote{Though admittedly premature, some other
physical properties of such would-be particles are fun to
contemplate: being classically stable they could easily have
cosmological life times. They would have a mass in the $\sim 10\  TeV$
region, zero charge and dipole moments in their ground state,
and geometrical interaction cross sections of order $1/m_W^2$.
Assuming maximum production at the electroweak phase transition,
a rough estimate of their present abundance shows that they
could be candidates for cold dark matter in the 
universe.}$\ $.
Mathematically the situation is the same as in the  
hidden-gauge-boson models\cite{Kunzz,Dobado} of strong 
 and electroweak interactions,
except that the role of the hidden gauge bosons is here played
by  $W^{\pm}$ and $Z$ themselves.


We carried out a numerical search\cite{Kunzz,Peccei,Tinyakov} 
for such stable spherically symmetric particle-like
soliton solutions in a simplified version of 
the two-Higgs $\rm SU(2)\times U(1)$ model.
The presumably inessential simplification consists 
of taking (a) the $\rm U(1)$ gauge
coupling $g^\prime=0$ as required
by spherical symmetry and (b) the Higgs potential to be the sum
of two "mexican hats" one for each Higgs doublet.
The resulting action is:   
\begin{equation}
  S={1\over {g^2}}\int d^4x \Bigl( -\half Tr(W_{\mu\nu}W^{\mu\nu}) 
  +\sum_{I=1,2} (D_{\mu} H_I)^{\dagger}(D^{\mu} H_I)
  +\sum_{I=1,2} {{\lambda_I}\over {g^2}} 
  (H^{\dagger}_I H_I- g^2 v^2_I)^2 \Bigr)
\end{equation}
with $ D_\mu H_I=(\partial_\mu + W_\mu) H_I ,\;
I=1,2 $ 
the covariant derivative on
the Higgs doublets. $W_\mu \equiv {1\over{2i}} \tau^a W^a_\mu$ and 
$W_{\mu\nu}$ is the ${\rm SU(2)}$ field strength.
We used the spherically symmetric
ansatz
  \[
  W_0=a_0\tau_in_i/2i\; ,
  \] 
  \[
  W_i=[ (\alpha-1)\epsilon_{ijk}\tau_jn_k/r + 
      \beta(\delta_{ij}-n_in_j)\tau_j/r + a_1n_in_j\tau_j ]/2i
  \]
  \[
  H_I=(\mu_I+i\nu_I n_i\tau_i)\xi
  \]
where $n_i$ is the unit vector in the direction of $\bf x$, $\tau_i$ are the
three Pauli matrices, $\xi$ is a constant unit doublet and $a_0, a_1,
\alpha, \beta, \mu_I, \nu_I$ are functions of $r$ to be determined
by the energy extremization. It is convenient to choose the
gauge $a_0=0$. One then solves Gauss' constraint
to obtain $a_1=0$\cite{Yaffe}. 

We rescale distances and fields by the appropriate 
powers of $m_W=g \sqrt{(v_1^2+v_2^2)/2}$ and use the notation
$\phi_I=\mu_I+i\nu_I \equiv F_I \e^{i\Theta_I}$ to write for the
energy functional:
  \[
  E = {M_W \over \alpha_W} 
  {2\over 1+t^2} \int dr \Bigl\{
  {1+t^2\over 2}(\a'^2+\b'^2) 
  + r^2(F_1'^2+F_2'^2+F_1^2\t_1'^2+F_2^2\t_2'^2)
  \] \nopagebreak
  \[
  + {1+t^2\over 4r^2}(\a^2+\b^2-1)^2
  +\half (\a^2+\b^2+1)(F_1^2+F_2^2) 
  \] \nopagebreak
  \[
  -\a[F_1^2\cos(2\t_1)+F_2^2\cos(2\t_2)]
  -\b[F_1^2\sin(2\t_1)+F_2^2\sin(2\t_2)]
  \] \nopagebreak
  \begin{equation}
  +{\k_1^2\over 4}r^2(F_1^2-1)^2
  +{\k_2^2\over 4t^2}r^2(F_2^2-t^2)^2 \Bigr\},
  \label{E'}
  \end{equation}
where $t\equiv {v_2}/{v_1}$ and $\k_I\equiv {m_{H_I}}/m_W$. The
radial coordinate in particular in the above expression
is measured in units of $m_W^{-1}$.

We require finiteness of the energy of the solution and use
the global symmetries of the energy functional\footnote{The energy
functional is invariant under the simultaneous rotation of the
complex fields $\phi_1, \phi_2 \;{\rm and}\; \alpha+i \beta$ by an angle
$\omega, \omega \;{\rm and}\; 2 \omega$ respectively, as well as
under independent rotations of $\phi_I$ by $\pi$.}$\;$ to specify the 
boundary conditions at $\infty$ and write
them in the form:
  \[
  \a\to 1, \;\;\;\;
  \b\to 0, \;\;\;\;
  F_1\to 1, \;\;\;\;
  F_2\to t, \;\;\;\;
  \t_a\to 0
  \]
Correspondingly, at $r=0$ energy finiteness implies 
  \[
  \a^2+\b^2\to 1.
  \]
while the field equations obtained by varrying (3) 
with respect to $\Theta_I$, together with the requirement
of smoothness of the solution lead to the dynamical
conditions: 
$
2 \Theta_a - \psi = 0 \;{\rm mod} \;2 \pi
$
, where $\psi$ is the phase of the 
field $\chi \equiv \alpha+i \beta$. 
Thus all solutions satisfy $  \Theta_1 -  \Theta_2 =  \pi N$
at the origin.
 
We have looked for solutions in a wide region of the
$\{t, \kappa_1, \kappa_2 \}$ parameter space. 
We refer the reader to [31] for
the detailed presentation of the numerical method we used and of 
the entire solution-zoo we have obtained so far. 
In the present note, I will 
describe briefly some of the solutions we found 
for various values of $\kappa$ on
the line $\{ t=1, \kappa_1=\kappa_2 \equiv \kappa \}$,
together
with their main characteristics.
As we varry $\kappa$ the energy landscape changes smoothly
and consequently, one generically 
expects a continuous change in the number and 
properties of its extrema and
saddles.
Figure 1 below captures these changes in part of the energy
landscape and shows some of the solutions of the model at hand. 

\hbox{
\hspace{0.57cm}
\psfig{file=graphtree.ps,bbllx=50bp,bblly=90bp,bburx=500bp,bbury=790bp,
height=8cm,width=15cm,angle=270}
}

\vspace {1.5cm}

The graph should be thought of as an "artist's conception" of a part 
of the tree
of solutions with $N=0$ and $|N|=1$ which has emmerged so far 
in the two-Higgs
model under study. Each branch of the "tree" corresponds to
a particular type of solution. As we move along a branch by varrying
$\kappa$, continuous quantities like the detailed
form of the solution, its energy as well as the value(s) of the
imaginary eigen-frequencies of the small oscillations around
it change smoothly, while integer-valued ones
such as the index $N$, the number of unstable negative curvature modes 
(shown on the branch) and 
the number of nodes of the fields, remain invariant.
We have suppressed the branches corresponding to the  
CP-conjugates of the solutions shown. They may be
imagined as the mirror images with respect to the
$\kappa-$axis of the branches above it.

For $\kappa$ between 0.0 and 5.5 the only non-trivial solution is 
the sphaleron embedded in the two-Higgs model;
it has $\Theta_1=\Theta_2=0$, $\beta=0$
and one negative mode. For $\kappa$ about 5.5 we reach the 
bifurcation point ${\rm B}_0$, the sphaleron
develops a second negative mode and at the same time two new solutions, 
namely the
branch $A_1$ and its CP conjugate 
appear with $N=1$ and $N=-1$ respectively.
The phases $\Theta_1$ and $\Theta_2$ and the complex field
$\chi$ of the solution corresponding to the branch $A_1$ 
have the behaviour we call 
of "type-A" and show in figure 2.  

\vspace {1.cm}
\hbox{
\hspace {0.7cm}
\psfig{file=graphTypeAS.ps,bbllx=50bp,bblly=260bp,bburx=500bp,bbury=500bp,height=6.cm,width=11.5cm,angle=0}

}

We have checked that the branch ${\rm A}_1$ extends without 
a new bifurcation
up to $\kappa=200$. As for the sphaleron, once we get to 
$\kappa \simeq 12.0$ we encounter the bifurcation ${\rm B}_1$. 
The sphaleron aquires a third unstable mode 
and two new solutions (branch ${\rm S}_2$ and its CP-conjugate) emerge.
They have $N=0$ and phase behaviour of "type-S" shown above.
And so on for the rest of the tree.
 
Incidentally, using the number of negative modes $N_{\rm neg. \;modes}$
of the solutions
shown in figure 1 one may explicitly verify 
the sum rule:
$
\sum_{\rm solutions} (-)^{N_{\rm neg. \;modes}} = 
{\rm constant}  
$
independent of $\kappa$,
for the whole range of values we considered.

\section{Discussion}

It is clear that the two-Higgs system discussed here posesses
a variety of unstable solutions, 
whose richness
increases with the Higgs masses. 
These, like the ordinary sphaleron, are in principle interesting in their
own right especially in connection with the baryon number generation
at the electroweak phase transition. 
There are important differences\cite{Tinyakov}  
in the values of the energies
and of the negative modes of the above solutions
compared to analogous ones 
of the minimal standard
model.
For instance, the value of the negative
curvature along the $A_1$
branch is about half of what it is
in the corresponding least unstable deformed 
sphaleron of the one-Higgs model. These differences 
affect directly the
prediction for the baryon number in the Universe, and at the same time 
they lay support to our general arguments
that an extended Higgs sector improves 
the stability of the winding solitons under discussion.

Despite of this evidence though, we have not as yet
been able to find a stable soliton.
Such a solution might for example arise at 
a bifurcation point like X of figure 1
and either merge with another branch of 
the tree at some new bifurcation
or continue, as is the case with $A_1$, up to $\kappa=\infty$.
But since our numerical method requires an initial
guess for the solution, which has to lie inside the basin of attraction 
of the corresponding
local minimum of the energy functional,
our failure to find a stable solution might just be 
due to the bad starting configurations we tried so far. Further
theoretical analysis is required to guide our numerical search.

\vspace{1cm}
{\bf Acknowledgements}
\vskip 0.4cm  

This research was supported in part by
the EEC grants CHRX-CT94-0621 and CHRX-CT93-0340,
as well as by the Greek General Secretariat
of Research and Technology
grant 91$E\Delta$358.

\vskip  0.6cm

\end{document}